%% file: generalisedBM_arxiv.tex
\documentclass{article}                     

\usepackage{graphicx}
\usepackage{enumerate}
\usepackage[utf8]{inputenc}

\usepackage{natbib}
\usepackage{hyperref}
\usepackage{setspace}
\usepackage{amsmath,amssymb,url}
 \usepackage{dsfont}
 \usepackage[normalem]{ulem}
\usepackage{multirow}
\usepackage{tikz}
\usetikzlibrary{positioning,shapes,arrows}
\usepackage{subcaption}

\DeclareMathOperator*{\argmax}{arg\,max}
\DeclareMathOperator*{\argmin}{arg\,min}

\newcommand{\bX}{\boldsymbol{X}}

\newcommand{\Xall}{\bX}
\newcommand{\Zall}{\bZ}
\newcommand{\ind}{\mathds{1}}
\newcommand{\Expec}{ \mathbb{E}}
\newcommand{\Prob}{ \mathbb{P}}
\newcommand{\prior}{ \pi}

\newcommand{\bZ}{\boldsymbol{Z}}
\newcommand{\bpi}{\boldsymbol{\pi}}
\newcommand{\balpha}{\boldsymbol{\alpha}}
\newcommand{\btau}{\boldsymbol{\tau}}
\newcommand{\bK}{\boldsymbol{K}}

\newcommand{\Ecal}{\mathcal{E}}
\newcommand{\Pcal}{\mathcal{P}}
\newcommand{\Ical}{\mathcal{I}}

\newcommand{\Acal}{\mathcal{A}}

\newcommand{\Scal}{\mathcal{S}}
\newcommand{\KL}{\mathrm{KL}}
\newcommand{\ICL}{\mathrm{ICL}}
\newcommand{\pen}{\mathrm{pen}}
\newcommand{\Rapprox}{\mathcal{R}_{\btau,\bX}}
\newcommand{\M}{\mathcal{M}_{\bK}}

\newtheorem{rem}{Remark}

\def\mathsmall#1{\mbox{\footnotesize $#1$}}

\newcommand{\HRule}[1][\medskipamount]{\par
  \vspace*{\dimexpr-\parskip-\baselineskip+#1}
  \noindent\rule{\linewidth}{0.2mm}\par
  \vspace*{\dimexpr-\parskip-.5\baselineskip+#1}}

\usepackage{xr}

\usepackage[affil-it]{authblk}

\begin{document}

\title{Block models for multipartite  networks\\ Applications in ecology and ethnobiology}


\author{Avner Bar-Hen  
\thanks{avner@cnam.fr}}
\affil{CNAM, 75003, Paris, France }
\author{P. Barbillon \thanks{pierre.barbillon@agroparistech.fr}}
\affil{ UMR MIA-Paris, AgroParisTech, INRA, Université Paris-Saclay, 75005, Paris, France}
\author{S. Donnet \thanks{sophie.donnet@inrae.fr}}
\affil{ UMR MIA-Paris, AgroParisTech, INRA, Université Paris-Saclay, 75005, Paris, France}

\providecommand{\keywords}[1]{\textbf{\textit{Keywords}} #1}

\date{\today}

\maketitle

\begin{abstract}
Generalized multipartite networks  consist in the joint observation of several networks implying some common individuals .
Such complex networks arise commonly in social sciences, biology, ecology, etc.
We propose a generative  probabilistic model named Multipartite Blocks Model (MBM) able to  unravel the topology of multipartite networks by identify clusters (blocks) of nodes sharing the same patterns of connectivity across the collection of networks they are involved in.  
The model parameters   are estimated  through a variational version of the  Expectation-Maximization algorithm.  The numbers of  blocks  are chosen  with an  Integrated Completed Likelihood criterion specifically designed for  our model. A simulation study illustrates the robustness of the inference strategy. Finally,  two datasets respectively issued from ecology and ethnobiology  are analyzed with  the MBM in order to illustrate its flexibility and its relevance for the analysis of real datasets.

The inference procedure is implemented in an \textsf{R}-package \textsf{GREMLIN}, available on Github   (\href{https://github.com/Demiperimetre/GREMLIN}{https://github.com/Demiperimetre/GREMLIN}). Simulations are  implemented as vignette of the package in order to ensure reproducibility of the results. 
\end{abstract}


\section{Introduction}
Networks have become fundamental tools in  various fields, such as biology, ecological theory or sociology  to name but a few.
Statistical analysis aims to study the structure of these networks for instance by unraveling the clusters or communities of individuals (nodes) shaping the observed interactions \citep[see][for a review]{Kolaczyk:2009}.  \\
The recent years have witnessed a growing interest for multi-layer networks which is a classical but wide terminology for a collection of related networks  \citep[see][for a review of   multi-layer networks in ecology for instance]{Pilosof2016}.
Among others, multi-layer networks  encompass dynamic networks, i.e. networks evolving with time \citep[see][for a review]{kim2018review}, multiplex networks when several types of relations are simultaneously studied on a common set of individuals \citep[see][for instance]{kefi,Barbillon2016d}  and multipartite networks.  \\
Multipartite networks are a generalization of bipartite networks.  In a bipartite network, the   nodes (representing the interacting entities) are partitioned into two disjoint sets and an edge links a node from one set to a node  from the other set (see Figure \ref{fig:schema bipar}, left). In a multipartite network, the nodes are divided into more than two sets and edges link entities from different sets (see   Figure \ref{fig:schema bipar}, middle).
In what follows, these pre-specified sets of nodes  will be referred to as  \emph{functional groups}.
Such multipartite networks arise in ecology when studying the interactions between several groups of species such as the interactions plant/pollinator, plants/ants, etc \cite[][]{Pocock_etal_12,Dattilo} or in biology when analyzing networks  issued from multi-omics datasets involving proteins, etc. \citep[see][for instance]{PAV18}.  \\
Generalized multipartite networks are an extension of  multipartite networks:  the nodes  are still partitioned into functional groups but the   interactions may occur not only between different functional groups but also  inside some of the functional groups.
Ethnobiology (which is  the scientific study of the relations between environment and people) provides  such networks. One of the problematic of ethnobiology  is to understand how social relations between individuals (seed circulation in our example)
may structure  and guarantee biodiversity in the cultivated crop species \citep[see][for example]{thomasExchanges}.  As a consequence,  the network of interest has two types of nodes --namely farmers and crop species-- and we observe interactions between farmers and crop species (who grows what) and inside the group of farmers (seed exchange). 
Marketing also provides generalized multipartite networks
when individuals are connected through social networks and also interact with goods through their on-line purchases. \\
\input{figure_tikZ.tex}

%
Many statistical tools have been proposed to analyze simple or bipartite networks.  However,  when individuals or biological entities are involved in several networks, there is a strong need to propose statistical tools handling these networks jointly. These tools are an essential step towards  a better understanding of the global interaction systems at stake.  The aim of this paper is to propose a probabilistic model suited for generalized multipartite networks.

A central issue in network analysis  is to be able to cluster nodes sharing the same connectivity patterns.  To do so, two approaches are possible,   either  relying on classical metrics detecting  nestedness \citep{almeida2008consistent},  modularity \citep{barber2007modularity}, etc, or adopting a probabilistic mixture strategy.  Since  the original paper by \cite{Snijdersnowicki1997}, probabilistic mixture models - classically referred as Stochastic Block models (SBM)-- have proven their efficiency when aiming to cluster  similar nodes  based on their connectivity patterns without any a priori hypothesis about the patterns to be found (e.g. modularity, centrality, hierarchy). Latent Block Models (LBM) \citep{Govaert2003}  are the extension of SBM for bipartite networks, resulting into a bi-clustering of the nodes.

In this paper, we propose a new  probabilistic mixture  model adapted to  generalized multipartite networks.  This model assumes that each functional group  is partitioned into clusters gathering nodes sharing the same connection behavior
on the basis of all the networks they are involved in.  The clusterings are introduced through latent variables resulting into a mixture model on edges.

Inferring the parameters of such a model with latent variables is a complex task, since the likelihood can not be computed in a close form.  We resort to a Variational Expectation-Maximization (VEM) algorithm to estimate the parameters and supplying the multi-clustering of the nodes.  The crucial task of estimating the numbers of clusters in each
functional group is tackled though a model selection procedure   based on  an adapted penalized likelihood criterion namely Integrated Classification Likelihood (ICL). 

\textbf{Related works}
Some extensions of standard descriptive  tools --such as community detection-- to more complex networks have been proposed in the literature  \citep[see for instance][]{yang2012community,Gasko:2017}. 
These metrics or descriptive tools have the drawback to only look for specific types of structures,  contrary to clustering methods based on probabilistic mixture  models. To the best of our knowledge, no such a probabilistic model has been proposed for generalized  multipartite network. However,
latent block models do exist for other kind of multi-layer networks.
When the network is multiplex (i.e. when several types of interactions are studied on the same set of nodes),  the SBM has been extended in \citet{kefi,Barbillon2016d}
by assuming a multivariate Bernoulli distribution to model the set of multiple edges between two individuals.
Extending the SBM to a dynamic context where several snapshots of the networks at different time points are available, was proposed in numerous papers. A main difference between these papers is whether the evolution is considered as been discrete \citep{xu2014dynamic,matiasmiele2017} or continuous \citep{dubois2013stochastic,xin2017continuous} in time.

\textbf{Outline of the paper}
 Section \ref{sec:data} is dedicated to the introduction of notations supplying a flexible tool to
describe generalized multipartite networks. We also provide a description of the two datasets of interest  and illustrate the  notations on these specific cases.
The  block model for multipartite networks  is described in Section  \ref{sec:mod}.  In Section~\ref{sec:inference}, the variational inference the model selection procedure are presented including the practical implementation of the algorithm.  Numerical illustrations of the robustness of the inference method are provided in Section \ref{sec:simu}.
 Finally, the statistical analyses of the  two datasets with discussion are presented in Section \ref{sec:real}. Perspectives are discussed in the last section.  Note  that this document is accompanied with a appendix sections  divided into three parts:  in the two first sections, the derivations of the update formulas for the VEM algorithm of the ICL criterion are provided;  finally, detailed estimates on the two datasets are given.  

\section{Notations and data}\label{sec:data}

A  generalized multipartite network can be seen as  a collection of  networks involving  $Q$ functional groups:  each network may be simple, i.e.  describing the relations inside  a given    functional   groups   or bipartite, i.e.  describing the relations between individuals belonging to two different functional groups.  Let $n_q$ be  the number of individuals  in the $q$-th functional group ($q=1,\ldots,Q$).
We index the collection of networks by pairs of functional groups $(q,q')$ ($q $ and $q'$ in $ \{1,\ldots,Q\}$).
The set $\mathcal{E}$ denotes  the  list of pairs of  functional groups  for   which we observe an  interaction network.
\\
For any $(q,q') \in \Ecal$, the interaction network is encoded in a matrix  $X^{qq'}$ such that  $X^{qq'}_{ii'} \neq 0 $ if there is an edge from unit $i$ of functional group
$q$ to unit $i'$ of functional group $q'$, $0$ otherwise.
Each network may be binary ($X^{qq'}_{ii'} \in \{0,1\}$) or valued ($X^{qq'}_{ii'}    \in \mathbb{N} $ or $\mathbb{R}$). 
 $X^{qq}$ may be  symmetric if the relation  is non-oriented, non-symmetric otherwise.
$\bX =\left(X^{qq'}\right)_{(q,q')\in \mathcal{E}}$ encodes the generalized  multipartite network.
For each network, $\Scal^{qq'}$ is an additional notation which refers to the list  of all the possible interactions.

\textbf{Illustration}
 $\bullet$ The  \textbf{dataset 1} is issued from \cite{Dattilo}. This ecological network gathers   mutualistic relations  between plants and pollinators, plants and ants,  and plants and frugivorous birds, resulting into $Q=4$ functional groups, namely  plants ($q=1$), pollinators ($q=2$), ants ($q=3$) and birds ($q=4$ and $\mathcal{E}=\{(1,2),(1,3),(1,4)\}$.  $X_{ii'}^{1q'}=1$ if the plant species $i$ has been observed at least once in a mutualistic interaction with the animal species $i'$ of
 functional group $q'$ during the observation period, $0$ otherwise.
 Data set is available at \url{https://figshare.com/articles/Interaction_matrix_A_in_which_elements_aij_1_represent_the_presence_of_an_interaction_between_the_plant_species_i_and_the_animal_species_j_and_zero_for_no_observed_interaction/3443210/2}.

$\bullet$   The \textbf{dataset 2}  comes from  \cite{thomasExchanges} and \cite{thomasInventory}. They collected the oriented network of seed circulation between farmers --resulting in a non-symmetric adjacency matrix -- and
 the  crop species  grown by the farmers, resulting in an incidence matrix. Noting $q=1$  for  the  farmers and $q=2$ for crop species we get  $\Ecal=\{(1,1),(1,2)\}$. $X^{11}_{ii'}=1$
 if farmer $i$ gives seeds to farmer $i'$ (oriented relation), $0$ otherwise and $X^{12}_{ij}=1$ if farmer $i$ cultivates crop species $j$, 0 otherwise.


\section{A block model for generalized multipartite networks}\label{sec:mod}

In order to account for heterogeneity among individuals, we propose a mixture model explicitly describing the way edges connect nodes in the various networks. 
\\
We assume that \emph{each functional group $q$  is  divided into $K_q$  blocks} (or equivalently clusters).  $\forall q\in \{1,\ldots,Q\}$ and  $ \forall i \in \{1,\ldots,n_q\}$, let $Z^{q}_i$ be
the latent random variable  such that $Z^ q_i =k$  if individual $i$ of functional group $q$ belongs to cluster $k$.
The random variables $Z^{q}_i$'s are assumed to be independent and  such that: $\forall k \in \{1,\ldots,K_q\}, \forall q \in \{1,\ldots,Q\}, \forall i \in \{1,\ldots,n_q\}$:
\begin{equation}\label{eq:mod2}
\Prob(Z^{q}_i=k) = \pi^{q}_k,
\end{equation}
with $\pi^{q}_k \in [0,1]$,  $\forall k= 1,\dots, K_q$ and $\sum_{k=1}^{K_q}\pi^{q}_k=1$,  $\forall q \in \{1,\ldots,Q\}$.
We set $\bZ^q =  \left(Z^{q}_i\right)_{i\in \{1,\ldots,n_q\}}$, $\bZ = \left(\bZ^q\right)_{q \in \{1,\ldots,Q\}}$ and $\bpi =\left(\pi^q_k\right)_{k \in \{1,\ldots, K_q\}, q \in \{1,\ldots,Q\}}$.
\\Then, the nodes connect as follows:  for any $(q,q') \in \Ecal$,
$\forall (i,i')\in \Scal^ {qq'}$,
\begin{equation}\label{eq:mod1}
X^{qq'}_{ii'} | \{Z^{q}_i=k, Z^{q'}_{i'}=k'  \} \sim_{ind} \mathcal{F}_{qq'}(\alpha^{qq'}_{kk'}).
\end{equation}
 $\mathcal{F}_{qq'}$ depends on the relation represented in   $X^{qq'}$. We assume that for any $(q,q')$,  $\mathcal{F}_{qq'}$ is either  the Bernoulli distribution if  the relation in $X^{qq'}$ is binary,  or the Poisson  distribution if $X^{qq'}_{ij}$ is a counting or the  Gaussian distribution if $X^{qq'}$ encodes a continuous strenght of interaction.    Equations \eqref{eq:mod2} and \eqref{eq:mod1} define the so-called Multipartite Block Model (MBM).

\begin{rem}\label{rem:depen}
  Our model is a generalization of the  SBM and  the LBM.  Indeed, Equations \eqref{eq:mod2}-\eqref{eq:mod1}
reduce to the SBM  if  $ \Ecal = \{(1,1)\}$ and to the LBM  if $ \Ecal = \{(1,2)\}$.
 Our extension assumes  that the  latent structures   $\mathbf{Z}$ are shared  among the  $X^{qq'}$ i.e. if a functional group $q$ is at stake in several $X^{qq'}$,  the same  $\mathbf{Z}^q$ impacts the distributions of the corresponding interaction matrices. In other words, the clusters gather individuals sharing the same properties of connection in  the collection of   networks.
Obviously, if each functional group appears in only one element of $\Ecal$, the MBM reduces to independent SBMs or LBMs.
 In terms of probabilistic dependence,  conditionally  on $\bZ$,
the    $\left(X_{ii'}^ {qq'}\right)$ are independent. However, $\bZ$ being latent,
their  marginalization introduces dependence  not only  between    the $X^{qq'}_{ii'}$ but also
 between the  matrices  $\left(X^{qq'}\right)_{(q,q' ) \in \Ecal}$. As a consequence,  the clustering  variables  $(Z^q_i)_{q \in \{1,\ldots,Q\}, i\ \in\{1,\ldots,n_q\}}$ are dependent once conditioned by  $\bX$. 
\end{rem}

For a given vector $\bK = (K_1, \dots, K_Q)$,  let $\theta_{\bK} =  (\balpha,\bpi)$ be the unknown parameters of the MBM  where
$\balpha =   \left(\alpha^{qq'}_{kk'}\right)_{(k,k') \in \Acal^{qq'}, (q,q')\in \Ecal}  $ are the connection parameters, with $\alpha^{qq'}_{kk'} \in \Gamma^{qq'} \subset \mathbb{R}^{d_{qq'}}$.
We have to perform two inference tasks: first, for a given vector $\bK$,   estimating $\theta_{\bK} $ and  $\bZ$, second, selecting the right  $\bK$.

\section{Parameter inference and model selection}\label{sec:inference}

Let $\ell(\bX,\bZ; \theta_{\bK}) $ denote the complete likelihood of the observations $\bX$ and the latent variables $\bZ$ for parameter $\theta_{\bK}$.  Equations \eqref{eq:mod2} and \eqref{eq:mod1} lead to:
\begin{equation}\label{eq:lik}
\begin{array}{cll}
\log \ell_c(\bX,\bZ; \theta_{\bK}) &=&  \sum_{q=1}^Q \sum_{i = 1}^{n_q} \sum_{k = 1}^{K_q} \ind_{\{Z_i^q=k\}} \log( \pi^q_k) \\
&&+  \displaystyle \sum_{(q,q') \in \Ecal} \sum_{(i,i') \in \Scal^{qq'} } \sum_{(k,k') \in \Acal^{qq'}} \ind_{\{Z^{q}_i=k, Z^{q'}_{i'}=k'\}}  f_{qq'}(X_{ii'}^{qq'},\alpha_{kk'}^{qq'})  \\
\end{array}
\end{equation}
where $f_{qq'}$ is the log-density of $\mathcal{F}_{qq'}$.
If $q \neq q'$,   $\Acal^{qq'} =  \{1,\ldots,K_q\}  \times \{1,\ldots,K_{q'}\} $. If the interaction are oriented, then $\Acal^{qq} = \{1,\ldots,K_q\} ^2$; otherwise
  $\Acal^{qq'} =  \left\{(k,k') \in \{1,\ldots,K_q\}  |\; \;     k\leq k'\right\}^2$.
$\bZ$ being  latent variables,  the likelihood   $   \ell(\bX; \theta_{\bK}) $ is obtained by  integrating  $   \ell_c(\bX,\bZ; \theta_{\bK}) $  over all the possible values of $\bZ$ denoted $\boldsymbol{\mathcal{Z}}  =  \{ (z_i^q)_{i\in \{1,\dots, n_q\}, q\in \{1, \dots, Q\}}| z_i^q \in \{1, \dots, K_q\}\}$:
\begin{equation}\label{eq:likmarg}
\log \ell(\bX; \theta_{\bK}) =\log \sum_{\bZ \in \boldsymbol{\mathcal{Z}}} \ell_c(\bX,\bZ; \theta_{\bK}).
\end{equation}

The summation over $\boldsymbol{\mathcal{Z}}$ in Equation  \eqref{eq:likmarg}
becomes quickly computationally intractable when $K_q$'s and/or the $n_q$'s increase. Moreover, we are interested in inferring the clusterings $\bZ$.
\subsection{Variational EM algorithm}\label{subsec:vem}

  In such models with latent variables, the EM algorithm \citep{dempster77} is a standard tool to maximise the   likelihood, taking advantage of the simple form of  $\log \ell_c(\bX,\bZ; \theta_{\bK}) $.
However, when  conditioned by the observations $\bX$, the $(Z_i^q)$ are not independent (see Remark \ref{rem:depen}). As a consequence, the E step of the EM algorithm  --which consists in the integration of  $\log \ell_c(\bX,\bZ; \theta_{\bK})$  against $\Prob(\bZ | \bX; \theta_{\bK}')$-- has no explicit expression.
In the context of  the SBM and the LBM,  the  variational  version of the EM (VEM) algorithm has proved to be   a powerful tool for maximum likelihood inference    \citep[see][]{Govaert2008, Daudinetal2008}. In  the VEM, the problem of the E-step is tackled by replacing  $\Prob(\bZ | \bX; \theta_{\bK}')$ by an approximation $\mathcal{R}_{\Xall,\btau}$ sought among distributions enforcing independence between the $Z_{i}^q$'s:
$$\mathcal{R}_{\Xall,\btau}(\bZ) = \prod_{q=1}^Q\prod_{i=1}^{n_q} (\tau_{ik}^q)^{\ind_{Z_i^q = k}}, \quad \mbox{ where} \quad \tau_{ik}^q =\Prob_{\mathcal{R}_{\Xall,\btau}}(Z_{i}^q  = k).$$
At the  VE-step,   $\btau$   is chosen   such that the Kullback-Leibler divergence $\KL[\Rapprox, \Prob(\cdot | \bX; \theta_{\bK})]$ is minimized.
The M-step is :
$  \widehat{\theta}_{\bK} = \argmax_{\theta}  \Expec_{\Rapprox}\left[\log \ell_c(\bX, \bZ; \theta)\right].$
Iterating steps (VE) and (M) leads to the maximisation of a lower boud of the likelihood:
$
\Ical_{\theta}(\Rapprox) = \log   \ell(\bX ; \theta)  - \KL[\Rapprox, \Prob(\cdot | \bX; \theta)]$. \\
For the MBM,  at iteration $(t)$, the algorithm VEM is as follows:

\vspace{2em}

\HRule[0.1pt]
\vspace{1em}
VEM for MBM
\HRule

$\bullet$ \textbf{VE Step.} Find $(\tau_{ik}^q)_{i,k,q}$ solving
\begin{eqnarray*}\label{eq:syst2}
 \tau_{ik}^q  &\propto& \exp \Bigg \{ \log\pi_k^q  +   \left[ \sum_{q' \in \mathcal{E}_q}  \sum_{i'=1}^{n_{q'}}\sum_{k' = 1}^{K_q'} f_{qq'}(X_{ii'}^{qq'},\alpha_{kk'}^{qq'}) \tau_{i'k'}^{q'} \right] \nonumber \\
&&  +   \ind_{(q,q)\in{\mathcal E}}  \sum_{j  | (i,j) \in  \Scal^{qq} } \sum_{k'=1}^{K_q} f_{qq}(X_{ij}^{qq},\alpha_{kk'}^{qq}) \tau_{jk'}^{q}
+   \ind_{(q,q)\in{\mathcal E}}     \ind_{(i,i) \in \Scal^{qq}}   f_{qq}(X_{ii}^{qq},\alpha_{kk}^{qq}) \Bigg\}\,.\nonumber  \\
 \end{eqnarray*}
$\bullet$ \textbf{M Step.} Update of the parameters:
\begin{equation}\label{eq:VEM}
\pi_{k}^q = \frac{1}{n_q} \sum_{i=1}^{n_q}\tau^q_{ik}\,
\quad , \quad \alpha^{qq'}_{kk'} = \frac{\sum_{(i,i')\in \Scal^{qq'}} X^{qq'}_{ii'}\tau^q_{ik}\tau^{q'}_{i'k'}}{\sum_{(i,i')\in \Scal^{qq'}}   \tau^q_{ik}\tau^{q'}_{i'k'}}.
\end{equation}
\HRule
 \vspace{1em}
 \begin{rem} Formula \eqref{eq:VEM} on $\alpha_{kk'}^{qq'}$  is valid for the expectations of the Bernoulli, the Poisson and the Gaussian distributions.  If $\mathcal{F}_{qq'}$ is a Gaussian distribution of variance  $v_{kk'}^{qq'}$, then:
$$
v_{kk'}^{qq'} = \frac{\sum_{(i,i')\in \Scal^{qq'}}( X^{qq'}_{ii'})^2\tau^q_{ik}\tau^{q'}_{i'k'}}{\sum_{(i,i')\in \Scal^{qq'}}   \tau^q_{ik}\tau^{q'}_{i'k'}} - (\alpha^{qq'}_{kk'})^2.
$$ \end{rem}
The details of the VEM algorithm are provided in the  Section \ref{app:VEM}.   As for any EM-type algorithm,  its initialization is critical and will be discussed later.
$(\hat{\theta}_{\bK}$, $\hat{\btau})$ denotes the VEM estimates. A by-product of this algorithm is an approximation of the conditional distribution $\Prob(\bZ | \Xall; \hat{\theta}_{\bK})$  by  $\mathcal{R}_{\Xall,\hat{\btau}}$.  $\bZ$ is thus estimated by:
\begin{equation}\label{eq:Zhat}
\hat{Z}_i ^q = \argmax_{k \in \{1, \dots, K_q\}}  \hat{\tau}_{ik}^q.
\end{equation}
The consistency of the VEM estimates has been established  for the SBM by \cite{bickel2013} and the  LBM by \cite{brault2017} while  \cite{mariadassou2015} study the behavior of  $\Prob(\bZ | \Xall; \hat{\theta}_{\bK})$ for the same models.

\subsection{A penalized likelihood criterion}\label{subsec:modselec}

In practice, the number of  clusters $\bK = (K_1,\dots,K_Q)$ is unknown and should be estimated.  We adopt a model selection  strategy where a model    $\M$ refers to the MBM with $\bK$ clusters.
Among many classical model selection criterion such as AIC, BIC and their variants,   \cite{biernacki2000} proposed the  Integrated Classification Likelihood (ICL).
ICL has proven its capacity to outline the clustering structure in networks in \cite{Daudinetal2008} (for simple networks),  \cite{keribin2015} (for bipartite networks)  or \cite{mariadassou2010} for valued networks.  Its success comes from the fact that  when traditional model selection criterion  essentially involves a trade-off between goodness of fit and model complexity, ICL values not only goodness of fit but also clustering
sharpness. Following the same line, we propose a modification of  the ICL adapted to generalized multipartite networks:
 
\begin{equation}\label{eq:ICL}
\ICL(\M) =\log  \ell_c(\Xall,\hat{\bZ}; \hat \theta_{\bK})-  \pen(\M)
\end{equation}
 where $$
\pen(\M) =   \frac{1}{2}\left\{ \sum_{q=1}^{Q} (K_q-1)\log(n_q)   +\left(\sum_{(q,q')\in\Ecal}d_{qq'} |\Acal^{qq'}| \right)  \log \left( \sum_{(q,q')\in\mathcal{E}} |\Scal^{qq'}| \right)\right\}$$
 and $\hat{\bZ}$ has been defined in Equation \eqref{eq:Zhat}.  The better model is chosen as
$\hat{\bK} = \argmax_{\bK} \ICL(\M)$.
The penalization term  $\pen(\M)$ is made up of two parts:  $ \sum_{q=1}^{Q} (K_q-1)\log(n_q)$ corresponds to the clustering distribution and involves  the number of nodes, while  the other term depends on the size of $\alpha_{kk'}^{qq'}$ and the numbers of possible edges in each network $|\Scal^{qq'}|$.   The derivation of the ICL criterion is detailed in  the  Section \ref{app:proofICL}.

\subsection{Practical  algorithm}\label{subsec:algopractice}

The practical choice  of the model and the estimation of its parameters are computational intensive tasks.  Assume that    $K_q \in \{1, \dots, K_{q}^\star\}$, then, ideally, we should  compare $\prod_{q=1}^Q K_{q}^\star$ models through the ICL criterion.
For each model, the VEM algorithm  has to be run starting from  a large number of initialization points  chosen carefully (due to its sensitivity to the starting point), resulting in
an unreasonable computational cost.  Instead, we propose to adopt a  stepwise strategy, resulting in a faster exploration of the model space  combined with  efficient initializations
of the VEM algorithm.  The procedure we suggest is the following one.

 Given a current model  $\mathcal{M}^{(m)} = \mathcal{M}(K^{(m)}_1, \dotsc, K^{(m)}_Q)$, the $m$-th iteration  of the procedure  is written as follows.

\vspace{1em}

\HRule[0.1pt]
\textbf{Model selection strategy for MBM}
\HRule

\noindent
$\bullet$ \textbf{Split proposals.}  For any $q$  such that  $K^{(m)}_q<K_{q}^\star$, consider the model
$$\mathcal{M}^{(m+1)q}_{+} = \mathcal{M}(K^{(m)}_1, \dots, K^{(m)}_q+1,\dots, K^{(m)}_Q).$$
\begin{itemize}
\item[$\cdot$]  Propose $K^{(m)}_q$ initializations  by splitting  any of the $K_q^{(m)}$ current  clusters into two clusters.
\item[$\cdot$] From each of the   $K^{(m)}_q$  initialization points,  run the VEM algorithm and keep the better variational estimate $\hat{\theta}_{\mathcal{M}^{(m+1)q}_{+}}$. 
\item[$\cdot$]  Compute the corresponding $\ICL(\mathcal{M}^{(m+1)q}_{+})$ from formula \eqref{eq:ICL}.
\end{itemize}
$\bullet$ \textbf{Merge proposals.} For any $q$  such that  $K^{(m)}_q>1$, consider the model
$$\mathcal{M}^{(m+1)q}_{-} =  \mathcal{M}(K^{(m)}_1, \ldots, K^{(m)}_q-1,\ldots, K^{(m)}_Q).$$
\begin{itemize}
\item[$\cdot$] Propose $K^{(m)}_q(K^{(m)}_q-1)/2 $ initializations  by  merging any pairs of clusters among the  $K_q^{(m)}$ clusters.
\item[$\cdot$] From each initialization point,  run the VEM algorithm and  keep the better variational estimate $\hat{\theta}_{\mathcal{M}^{(m+1)q}_{-}}$
\item[$\cdot$] Compute the corresponding $\ICL(\mathcal{M}^{(m+1)q}_{-} )$.

\end{itemize}
$\bullet$ Set $\mathcal{M}^{(m+1)}  =  \displaystyle \argmax_{\mathbb{M}^{(m)}  } \ICL(\mathcal{M} )$ where
 $
\mathbb{M}^{(m)} =\{\mathcal{M}^{(m)}\}  \cup  \bigcup_{q \in \{1, \dots Q\}} \{\mathcal{M}^{(m+1)q}_{+} \} \cup  \{\mathcal{M}^{(m+1)q}_{-} \}$.

If $\mathcal{M}^{(m+1)}  \neq \mathcal{M}^{(m)}$ iterate, otherwise stop.

\HRule

\vspace{1em}


\noindent Initializing the VEM from the clusters obtained on a smaller or larger model is much more efficient  than other strategies such as random initialization or spectral clustering.  Note that the tasks at each iteration can be  parallelized.

\section{Illustration on simulated datasets}\label{sec:simu}
We  illustrate the efficiency of our inference procedure on simulated data.
 \input{simu_study}

\section{Applications}\label{sec:real}


\subsection{Ecology: interactions plants / animals}\label{subsec:real ecolo}

\textbf{Dataset}
 The dataset --compiled and conducted by   \cite{Dattilo} at Centro de Investigaciones Costeras La Mancha (CICOLMA), located in Mexico--
 involves three general types of plant-animal mutualistic interaction:
 pollination,  seed dispersal by frugivorous birds, and   protective mutualisms between ants and plants with extrafloral nectaries.
The dataset --which is one of the largest compiled so far with respect
to species richness, number of interactions and sampling effort--  includes  $n_1 =  141$  plant species,
 $n_2 = 173$  pollinator species, $n_3 = 46$ frugivorous bird species and $n_4 = 30$  ant species, inducing  $753$  observed interactions of which $55\%$ are   plant-pollinator interactions, $17\%$ are   plant-birds interactions  and $28\%$ are plant-ant interactions.

\begin{figure}[ht]
\centering
   \includegraphics[scale=0.45]{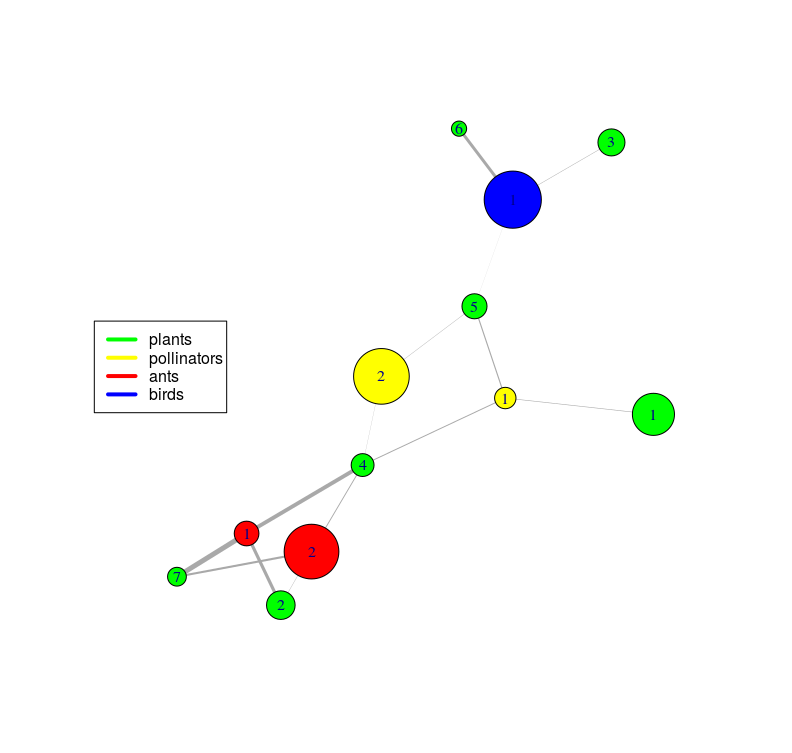}
   \caption{Mesoscopic view of the interconnected network. The size of the nodes are proportional to the size of the clusters
   and the width of the edges are proportional to the probability of connection between/within clusters. Edges corresponding to probabilities of connection lower than 0.01 are not plotted.}
   \label{fig:estim dattilo}

\end{figure}

\textbf{Inference}
 We run the procedure described in Subsection \ref{subsec:algopractice} starting from several automatically chosen  initial points $\bK^ {(0)}$, with numbers
 of clusters bounded between $1$ and $10$.    The computational time on an \emph{Intel   Xeon(R) CPU  E5-1650 v3 3.50GHz x12}  using $6$ cores is less than 10 minutes.
 The ICL criterion selects  $7$ clusters  of plants, $2$ clusters of pollinators, $1$ cluster  of birds and $2$  clusters  of ants.
The  estimated parameters are  reported in Tables \ref{tab : estim Dattilo}  and  \ref{tab : estim pi Dattilo} in the supplementary material. They are really similar to the ones used for the simulation study and provided in Equation \ref{eq:param sim}. 
The resulting mesoscopic view of the multipartite network is plotted in Figure \ref{fig:estim dattilo}.

\textbf{Discussion} From Figure \ref{fig:estim dattilo}, we  conclude  that  plants  of  Clusters $7$ and $2$ only interact with ants (plants of Cluster $7$ attract  more  ant species belonging to Cluster  $1$). The plants of Clusters $3$ and $6$ are only  in interaction   with birds, the difference between the two clusters being due to the strength of the connection.  The difference between the two clusters of pollinators derives from the existence of   Cluster $1$ of plants.

In order to illustrate the  contribution of our method, we   also analyze each bipartite network separately (using an LBM) and compare the results in terms of clustering.
The clusterings are compared through the Adjusted Rand Index (ARI):  if $ARI=1$ then the clusterings are equal (up to a label switching transformation).
The ARIs are given in Table \ref{tab:ARIdattilo}.

Using standard LBM, we obtain $2$ clusters of ants, $1$ cluster of birds and $3$ clusters of pollinators.
The clusterings of ants and birds are not modified by the ecosystemic approach, their ARI being equal to $1$. The clustering of the pollinators is slightly modified, going from $3$ clusters to $2$ clusters but the additional block only contains few individuals, thus leading to a high ARI.
Since the plants functional group is  involved in the three bipartite networks, we obtain $3$ clusterings when analyzing them separately.
These three clusterings are --as expected--  very different from our MBM clustering (the ARIs being  respectively equal to $0.118$, $0.415$ and $0.163$, see Table \ref{tab:ARIdattilo}).
When aiming at proposing a clustering taking into account the $3$ bipartite networks, one may adopt a naive strategy by  combining (by intersection) these three clusterings.
We then obtain $12$ blocks of plants and the ARI  with the MBM clustering is $0.617$. However, this number of clusters ($12$) is too large with respect to the model selection criterion.
Our MBM clustering is a parsimonious trade-off.

Finally,  Figure \ref{fig:estim dattilo} highlights the  central role played by the pollinators in the complete network also  involving ants and seed dispersal birds. Our ecosystemic approach unables us to unravel such a central position of pollinators in general with respect to ants and seed dispersal birds, which would have been impossible when performing separate  analysis.

\begin{table}[ht]
\centering
\caption{Comparison of clusterings when the networks are jointly modeled by the MBM  (denoted Full) and when the networks are considered apart as bipartite networks.
Inter denotes the clustering obtained by intersecting the three clusterings on plants for each bipartite network. The  selected number of clusters (in parenthesis)  and the ARIs  are provided.}
\label{tab:ARIdattilo}
\begin{tabular}{rrrrr}
  \hline
 & Full/Flovis & Full/Ants & Full/Birds & Full/Inter \\
  \hline
\multirow{ 2}{*}{Plants} & (7/3) & (7/3) & (7/3) & (7/12)\\
& 0.118 & 0.415 & 0.163 & 0.617 \\
\multirow{ 2}{*}{Flovis}& (2/3) &  &  &  \\
& 0.997 &  &  &  \\
\multirow{ 2}{*}{Ants}&  & (2/2) &  &  \\
&  & 1.000 &  &  \\
\multirow{ 2}{*}{Birds} &  &  & (1/1)&  \\
  &  &  & 1.000 &  \\
 \hline
\end{tabular}
\end{table}
\subsection{Seed circulation and crop species inventory}
\textbf{Dataset}
Seed circulation among farmers is a key process that shapes crop diversity \citep{coomes2015farmer,pautasso2013seed}.
Data on seed circulation  within a community of first-generation migrants ($30$ farmers)
were collected by a field survey in the island of Vanua Lava in the South Pacific archipelago nation of Vanuatu.
A farmer is considered as a giver for another farmer if he/she has given at least one crop since they arrived in the new settlement site in Vanua Lava. It results in a connected and directed network of seed circulation.
Besides the circulation network, inventory data for each farmer were also collected. They consist
in the list of crop landraces they grow. This was aggregated at the species level, leading to $37$ crop species. These inventory data can be seen as a bipartite network.
The seed circulation data were analyzed in \cite{thomasExchanges} and the inventory data were analyzed in a meta-analysis in \cite{thomasInventory}.
On the basis on the joint modeling we propose in this paper, we aim to provide a clustering on farmers and crop species on the basis of the seed circulation network
and the inventory bipartite network. 

\begin{figure}[h!]
 \centering
 \includegraphics[scale=.35]{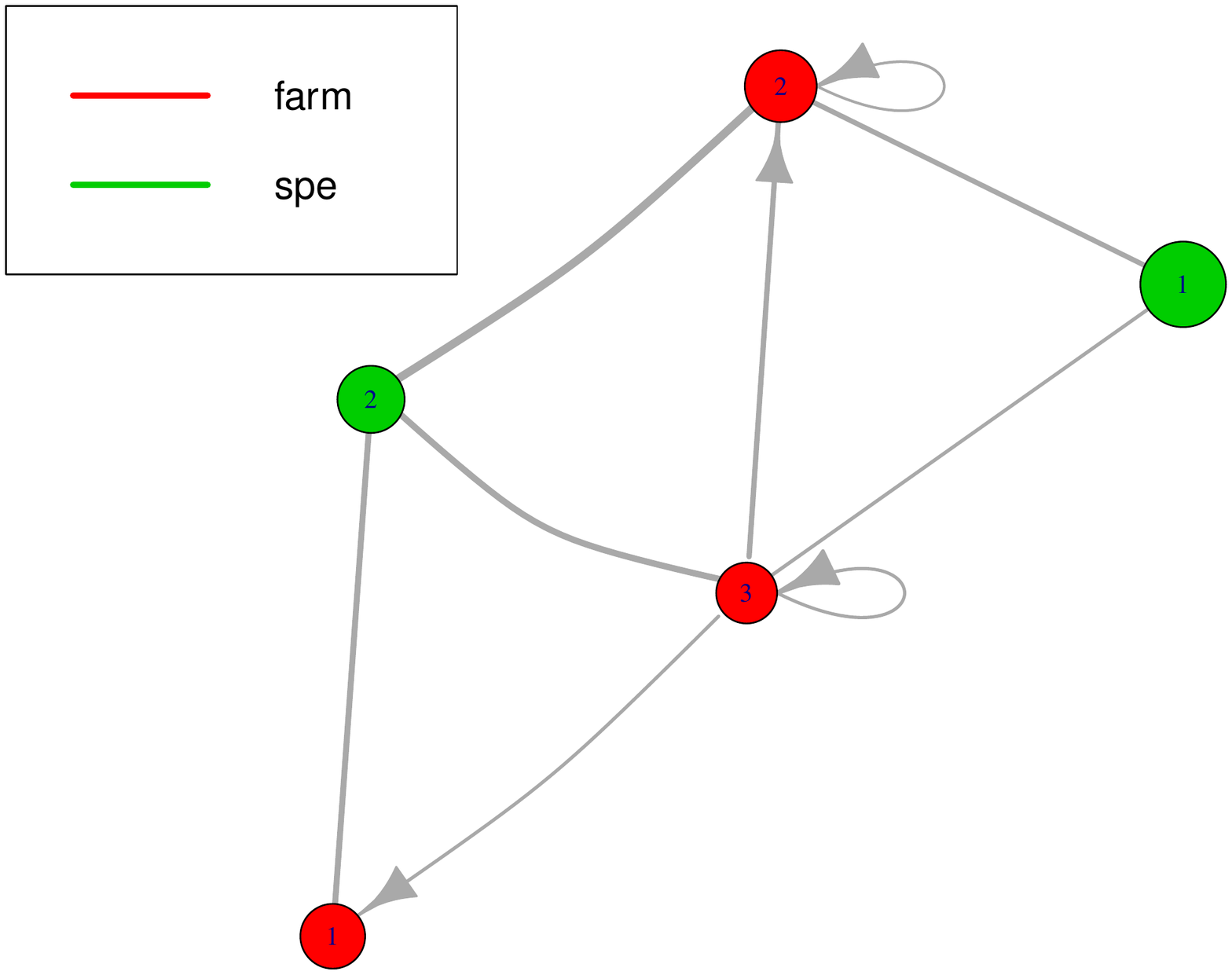}
 \caption{Summarized network provided by the MBM. Nodes correspond to the clusters detected by the MBM: clusters of farmers are in red and clusters of crop species are in green.
 Size of a node is proportional to the number of farmers or crop species belonging to the considered cluster.
The width of the edges are proportional to the probability of connection between/within clusters.
 The probability of connection below 0.2 are not plotted.}
 \label{figMIRES}
\end{figure}

\textbf{Inference}
Three clusters of farmers and two clusters of crop species were selected. The inferred parameters are given in Table \ref{tab:estimVanuatu} and a
a mesoscopic view  is displayed in Figure \ref{figMIRES}. The clusters were  renumbered to make them correspond to the probability of connection: the larger cluster number, the larger marginal
 probability of connection for the members of the cluster.

\begin{figure}
 \centering
 \caption{Alluvial plot comparing the clustering on farmers obtained from an SBM on the circulation network (clExc), an LBM on the inventory network (clInv) and the MBM (clMBM)
 on both networks. The cluster numbers are related with the probability of connection, the larger cluster number, the larger marginal probability of connection (between
 farmers for clExc, between farmers and crop species for clInv).}
 \label{figMIRESallu}
 \includegraphics[scale=.35]{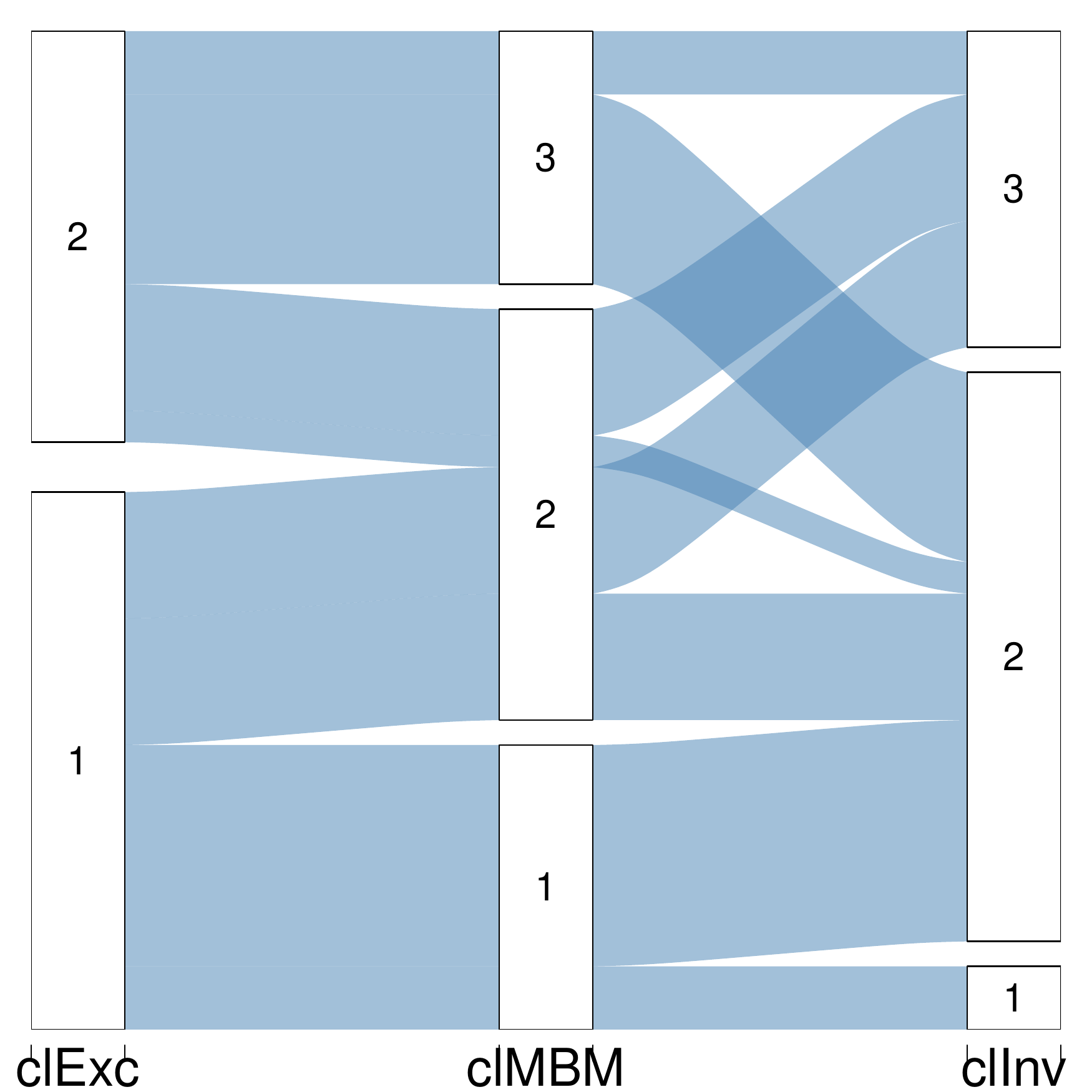}
\end{figure}

\textbf{Discussion}
The discovered clusters are straightforwardly interpretable: Cluster 3 gathers farmers who circulate seeds within the cluster and give to the two other clusters,
Cluster 2 circulates seeds within the cluster contrary to Cluster 1 who only receives from Cluster 3; the two clusters of crop species are Cluster 2
with more common crop species and Cluster 1 with
other species. Clusters 3 and 2 of farmers grow crop species from Clusters 1 and 2 whereas Cluster 1 of farmers grows mainly crop species from Cluster 2.
It turns out that Cluster 3 gathers mainly the first migrants and Cluster 1 the last migrants.
The pattern of connection is then explained by the fact that first migrants helped the others to settle by providing seeds. Moreover, the first migrants had more time
to collect more crop species to grow.
In order to compare the  clusterings obtained by the MBM and the ones obtained from the  circulation network  (clustering on farmers from SBM)  and  the inventory network (clusterings of farmers and plants with a LBM), we compute the ARI.
The clusterings on crop species remain quite close since their ARI is equal to $0.891$.
However, the clusterings on farmers are quite different (ARI smaller than $0.3$), indeed the MBM shall
make a trade off between the circulation and the inventory for farmers.
To ease the comparison between clusterings on farmers,
the same renumbering rule was applied for all the different clusterings so that the larger cluster numbers correspond to larger marginal probability of connection.
Figure \ref{figMIRESallu} is an alluvial plot which compares the three obtained clusterings of farmers.
It shows how the trade-off is made between the two stand-alone clusterings in the MBM clustering.
It appears quite obvious that Cluster 1 given by the MBM gathers only farmers from Cluster 1 in the seed circulation network and from Clusters 1 and 2 from the inventory data since
this cluster aggregates farmers with fewer connections and who grow less crop species than the others. The same kind of observation can be made for Cluster 3 given by the MBM
which aggregates farmers who were in the cluster of the most connected farmers and in the two clusters of farmers who grow more seed.

\section{Discussion}\label{Sec:disc}
In this paper, we proposed  an extension of LBM and SBM which can handle multipartite networks, resulting in the so-called MBM.
Multipartite networks encompass a lot of various situations such as the two examples dealt with in the paper. Besides,
MBM can be also useful for many other contexts with different multipartite structures.
The main limiting factor of the parameter inference and model selection methods we propose, is the size of networks. Indeed,
the inference algorithm is suitable for networks up to $1000$ nodes in order to keep computational time reasonable.
If willing at studying larger networks, one should develop adapted inference algorithms. \\
Several extensions can be thought of with no additional significant difficulty.
For instance, covariates can  be taken into account  by writing $\Prob(X_{ii'}^{qq'} = 1 | Z^q_i=k, Z_{i'}^{q'} = k')
= \phi\left(\alpha^{qq'}_{kk'} +\mathbf{y}^{qq'}_{ii'} \beta\right)$ where $\mathbf{y}^{qq'}_{ii'}$ are the covariates describing the pair of individuals $(i,i')$.
A more challenging extension could be to include extra parameters in the model to
account for degree heterogeneity in the vein of the degree-corrected SBM
\citep{karrer2011}. However, this should lead to cluster individuals independently of their degrees which is not always desirable. For instance in the seed circulation example, the in-degrees and out-degrees of farmers are key factors to explain their social structure. In the interaction networks in ecology, distinguishing generalist (high degree) from specialist (low degree) species is a part of the expected outcomes of the clusterings. More recent works \citep{zhu2014} propose to use node degrees and edge orientations to help the clustering while tolerating heavy-tailed distribution of degrees. Incorporating these variants may be of interest in the MBM framework and could lead to further developments.

In a more general perspective, the study of ecological or sociological interactions supplies a wide variety of complex networks
such as multilevel networks or multi-layer networks\citep{Pilosof2016,lazega2015}.
Some of them can be treated as multipartite networks and then by a MBM (possibly by incorporating some extensions discussed above). The others
 require the development of suited models which might also rely on a latent variable modeling. They will be the subject of future works.

\appendix

\input{annexe_detailsVEM}

\input{annexe_ICL}

\section{Detailed Estimates for the MBM on the two Datasets} \label{app:tables}
 \begin{table}[ht]
\centering
\begin{tabular}{|c|c|cc|cc|c|}
\hline
\hline
\multicolumn{2}{|c|}{} &  \multicolumn{2}{c|}{Pollinators} & \multicolumn{2}{c|}{Ants} & \multicolumn{1}{c|}{Birds}\\
 \cline{3-7}
 \multicolumn{2}{|c|}{$\hat{\alpha}^{qq'}_{kk'}$}  &1 & 2 & 1 & 2 & 1 \\
\hline
\parbox[t]{2mm}{\multirow{7}{*}{\rotatebox[origin=c]{90}{Plants}}} & 1  & 0.0957 & 0.0075 & 0 & 0.0006$^ *$ &0.0013$^*$ \\
 & 2 & 0.0042$^*$ & 0  & 0.5431 & 0.0589 & 0  \\
 & 3 & 0 & 0.0003$^*$ &  0 & 0   & 0.0753 \\
 & 4  & 0.1652 & 0.0343 & 0.6620 & 0.1542 &  0 \\
 & 5  & 0.1918 & 0.0638 & 0 & 0  & 0.0163 \\
 & 6 & 0 & 0 & 0  &0   & 0.5108 \\
 & 7  &0  &0  & 0.8492 & 0.3565 & 0  \\
 \hline
\hline
\end{tabular}
\caption{$\hat{\alpha}_{kk'}^{qq'}$ for  MBM on the plant/animals interactions dataset. Elements with $^*$ are not plotted in Figure \ref{fig:estim dattilo}.}
\label{tab : estim Dattilo}

\end{table}

\begin{table}[ht]
\centering
\begin{tabular}{|c|cc||c|cc|}
  \hline
\hline
\multirow{7}{*}{$\hat{\pi}^1$}  & 1&  0.4675 &\multirow{2}{*}{$\hat{\pi}^2$}&1 &0.06    \\
   & 2 &  0.1606  &&2 & 0.94  \\
\cline{4-6}
 &3& 0.1351 & \multirow{2}{*}{$\hat{\pi}^3$} &  1&  0.1\\
 &4 & 0.0784$^ *$  &  & 2  &   0.9\\
 \cline{4-6}
&5& 0.1061&  $\hat{\pi}^4$ & 1& 1 \\
\cline{4-6}
& 6 & 0.0142 & && \\
& 7 &  0.0380 & &&\\
  \hline
  \hline
\end{tabular}
\caption{$\hat{\pi}_{k}^{q}$ for  MBM on the plant/animals interactions dataset}
\label{tab : estim pi Dattilo}

\end{table}

 \begin{table}[ht]
\centering
\caption{Estimated parameters for  MBM on the seed circulation dataset: $\hat\pi_{k}^q$ are in the first row and column, other rows and columns contain the  estimates $\widehat{\alpha}_{kk'}^{qq'}$.
$\widehat{\alpha}_{kk'}^ {qq'}$ identified by $^*$ are not plotted in Figure \ref{figMIRES}.}
\label{tab:estimVanuatu}
\begin{tabular}{cccccccc}

\multicolumn{3}{c}{} &  \multicolumn{3}{c}{Farmers} & \multicolumn{2}{c}{Species} \\
\hline
\multicolumn{3}{c}{}  &1 & 2 & 3 & 1 & 2 \\
\hline

&  &  $\hat\pi_{k}^q$ & 0.31  & 0.42 & 0.27 & 0.65 &  0.35  \\
\hline
\parbox[t]{2mm}{\multirow{3}{*}{\rotatebox[origin=c]{90}{Farmers}}} & 1 & & 0.025$^*$ & 0.123$^*$ & 0.053$^*$ & 0.186 &0.653 \\
 & 2 & & 0.159$^*$ & 0.300  & 0.070$^*$ & 0.559 & 0.905  \\
 & 3 & & 0.374 & 0.585 &  0.357 & 0.390   & 0.696 \\
 \hline
\end{tabular}
\end{table}
\clearpage


\section*{Acknowledgements}

We thank Sophie Caillon (CEFE) for sharing the seed circulation and inventory data, Sophie Caillon and Mathieu Thomas (CIRAD) for related discussion concerning the analysis.
Support of the ongoing research collaboration through the MIRES network was provided by the Institut National de la Recherche Agronomique (INRA).
P. Barbillon and S. Donnet are funded by the ANR project Econet.

 \bibliographystyle{apalike}
\bibliography{biblio}

\end{document}

%% file: figure_tikZ.tex
\begin{figure}
\centering

\begin{minipage}[t]{0.2\linewidth}
\begin{tikzpicture}[
  mynoder/.style={draw,circle,text width=0.5cm,fill = red,align=center,scale=0.6},
  mynodeg/.style={draw,circle,text width=0.5cm,align=center, fill = green,scale=0.6},
 ]

\node[mynoder] (R1) {$1$};
\node[mynoder, below=0.1 cm of R1] (R2) {$2$};
 \node[mynoder, below=0.1 cm of R2] (R3) {$3$};
 \node[mynoder, below=0.1 cm of R3] (R4) {$4$};
 \node[mynoder, below=0.1 cm of R4] (R5) {$5$};
 \node[mynoder, below=0.1 cm of R5] (R6) {$6$};
 
\node[mynodeg, right =1 cm of R1] (G1) {$1$};
\node[mynodeg, below=0.1 cm of G1] (G2) {$2$};
 \node[mynodeg, below=0.1 cm of G2] (G3) {$3$};
 \node[mynodeg, below=0.1 cm of G3] (G4) {$4$};
 \node[mynodeg, below=0.1 cm of G4] (G5) {$5$};

\path[thick]
(R1) edge (G2)
(R2) edge (G3) 
(R2) edge (G4)
(R4) edge (G1)
(R4) edge (G5)
(R6) edge (G5);

\end{tikzpicture}

    \end{minipage}\hfill
\begin{minipage}[t]{0.5\linewidth}
\centering
\begin{tikzpicture}[
  mynoder/.style={draw,circle,text width=0.5cm,fill = red,align=center,scale=0.6},
  mynodeg/.style={draw,circle,text width=0.5cm,align=center, fill = green,scale=0.6},
  mynodeb/.style={draw,circle,text width=0.5cm,align=center, fill=cyan,scale=0.6},
mynodey/.style={draw,circle,text width=0.5cm,align=center, fill=yellow,scale=0.6}
]
\foreach \a in {1,2,...,5}{
\node[mynoder] at ({90/6 * (\a)}:  3cm) (R\a) {$\a$};
}
\foreach \a in {1,2,...,6}{
\node[mynodeb] at ({90 + 90/6 * (\a-1)}:  3cm) (B\a) {$\a$};
}

\node[mynodeg, below right  = 0.5 cm of B6] (G1) {$1$};
\node[mynodeg, right =0.5  cm of G1] (G2) {$2$};
 \node[mynodeg,right =0.5  cm of G2] (G3) {$3$};
 \node[mynodeg,right =0.5  cm of  G3] (G4) {$4$};
 \node[mynodeg, right =0.5  cm of G4] (G5) {$5$};

\node[mynodey, below = 0.5 cm of G1] (Y1) {$1$};
\node[mynodey, right =0.5  cm of Y1] (Y2) {$2$};
 \node[mynodey,right =0.5  cm of Y2] (Y3) {$3$};
 \node[mynodey,right =0.5  cm of  Y3] (Y4) {$4$};
 \node[mynodey, right =0.5  cm of Y4] (Y5) {$5$};
 \path[thick]
(R2) edge (G3) 
(R3) edge (G5) 
(R2) edge (G4)
(R4) edge (G1)
(R4) edge (G5);

\path[thick]
(G1) edge [dashed] (B1)
(G2) edge [dashed] (B4)
(G4) edge [dashed] (B4)
(G4) edge [dashed] (B5)
(G5) edge [dashed] (B3);

\path[thick]
(G1) edge (Y1)
(G2) edge (Y4) 
(G2) edge (Y3)
(G4) edge (Y2)
(G4) edge (Y1);

\end{tikzpicture}
    \end{minipage}\hfill
\begin{minipage}[t]{0.3\linewidth}
\centering
\begin{tikzpicture}[
  mynoder/.style={draw,circle,text width=0.5cm,fill = red,align=center,scale=0.6},
  mynodeg/.style={draw,circle,text width=0.5cm,align=center, fill = green,scale=0.6},
  mynodeb/.style={draw,circle,text width=0.5cm,align=center, fill=red,scale=0.6}
]
%
 
\node[mynodeg, below =1 cm of R1] (G1) {$1$};
\node[mynodeg, right=0.1 cm of G1] (G2) {$2$};
 \node[mynodeg, right = 0.1 cm of G2] (G3) {$3$};
 \node[mynodeg, right=0.1 cm of G3] (G4) {$4$};
 \node[mynodeg, right=0.1 cm of G4] (G5) {$5$};

\node[mynodeb, below =1 cm of G1] (B1) {$1$};
\node[mynodeb, right =0.1 cm of B1] (B2) {$2$};
 \node[mynodeb, right =0.1 cm of B2] (B3) {$3$};
 \node[mynodeb, right =0.1 cm of B3] (B4) {$4$};
 \node[mynodeb, right=0.1 cm of B4] (B5) {$5$};
 \node[mynodeb, right=0.1 cm of B5] (B6) {$6$};


\path[thick]
(G1) edge (B1)
(G2) edge (B4) 
(G2) edge (B2)
(G4) edge (B6)
(G4) edge (B3)
(G5) edge (B6);

\path[->] 
(B1) edge [bend right = 70] (B2)
(B3) edge [bend right = 70] (B5)
(B5) edge [bend left = 80] (B1)
(B6) edge [bend left = 70] (B4);
\end{tikzpicture}
    \end{minipage}

\caption{Illustrations of bipartite (left), multipartite (center) and generalized multipartite networks (right)}
\label{fig:schema bipar}
\end{figure}

%% file: simu_study.tex
\textbf{Simulation parameters}
We generate datasets mimicking the two real datasets studied in Section \ref{sec:real},
i.e. we generate $100$ datasets under the following two scenarios.

Simulations are  implemented as vignette of the R-package  available on Github   (\href{https://github.com/Demiperimetre/GREMLIN}{https://github.com/Demiperimetre/GREMLIN}) in order to ensure reproducibility of the results.

$\bullet$ \textbf{Scenario 1}: We set $Q =4$  functional groups of respective sizes  $n = (141,173,46,30)$, $\mathcal{E} = \{(1,2),(1,3),(1,4)\}$ and $K= (7,2,2,1)$. The simulation parameters are the following:
$\pi^1 =  (0.3651 , 0.1270 ,0.1190,0.1460 ,0.0842, 0.0794 ,0.0794 )$, $\pi^2 =(0.1,  0.9)$, $\pi^3=(0.1 , 0.9)$,
\begin{equation}\label{eq:param sim}
\alpha^{12} = \mathsmall{\left(
\begin{array}{cc}
0.0957 & 0.0075  \\
0.0100 & 0    \\
 0 & 0.0003  \\
 0.1652 & 0.0343  \\
0.2018 & 0.1380   \\
0 & 0 \\
0  &0  \\
 \end{array}
\right)}, \quad
\alpha^{13} =  \mathsmall{\left(
\begin{array}{cc}
 0 & 0.0006 \\
0.5431 & 0.0589  \\
 0 & 0 \\
   0.6620 & 0.1542   \\
   0 & 0   \\
 0  &0    \\
 0.8492 & 0.3565    \\
 \end{array}
\right)}   \mbox{, } \quad
\alpha^{14} = \mathsmall{ \left(
\begin{array}{c}
0.0013 \\
 0  \\
0.0753 \\
 0 \\
 0.0163 \\
0.5108 \\
 0  \\
 \end{array}
\right)}
.
\end{equation}

$\bullet$ \textbf{Scenario 2}: Mimicking our second dataset, we set $Q=2$ functional groups such that $n_1 = 30, n_2=37$, $\mathcal{E} = \{(1,1),(1,2)\}$ and $K= (3,2)$. The simulation parameters are:
$\pi^1 = (0.31,0.42,0.27)$, $\pi^2 = (0.65,0.35)$.
$$
\alpha^{11} =\mathsmall{\left(
\begin{array}{ccc}
 0.025 & 0.123& 0.053 \\
0.159& 0.3 &0.07\\
0.374 & 0.585&0.357
\end{array}\right)} \quad
\alpha^{12} =
\mathsmall{\left(
\begin{array}{cc}
0.186 & 0.653\\
0.559&0.905\\
0.390&0.696
\end{array}
\right)}.
$$

On each simulated dataset, we run the algorithm described in Section \ref{subsec:algopractice}  with   $K_{q}^\star = 10$, for all $ q$.   We start the stepwize algorithm on   $ K_q = 1$  and on the clusterings obtained by inferring  the LBM separately on each matrix $X^{qq'}$.
For each dataset, it takes  a few minutes  for the algorithm to converge.

\textbf{Results for Scenario 1}
\begin{table}
\centering
 \begin{tabular}{cccc|c}
 \multicolumn{4}{c }{$\hat \bK$} & \\
\hline
  $\hat K_1$ & $\hat K_2$ &$\hat K_3$  &  $\hat K_4$  & Nb of simulations    \\
  \hline
7 & 2 & 2 & 1 & 73 \\
   7 & 2 & 1 & 1 & 6 \\
    7 & 1 & 2 & 1 & 3 \\
  6  & 2 & 2 & 1 & 18 \\
   \hline\\
\end{tabular}
\caption{Results for  Scenario $1$:  estimated numbers of clusters $(\hat K_1, \hat K_2, \hat K_3, \hat K_4)$. The simulated number of clusters is equal to $(7,2,2,1)$.}
\label{tab : estim K Dattilo}
\end{table}

\begin{table}[ht]
\centering
\small
\begin{tabular}{|ll|ll||ll|ll||ll|}
  \hline
\multicolumn{4}{|c||}{$\alpha^{12}$} &  \multicolumn{4}{c||}{$\alpha^{13}$} & \multicolumn{2}{c|}{$\alpha^{13}$} \\
\hline
\hline
\multicolumn{2}{|c|}{$\alpha_{k1}^{12}$} &  \multicolumn{2}{c||}{$\alpha^{12}_{k2}$} &  \multicolumn{2}{c|}{$\alpha^{13}_{k1}$} &   \multicolumn{2}{c||}{$\alpha^{12}_{k2}$} &   \multicolumn{2}{c|}{$\alpha^{14}_{k1}$}\\
\hline
  \hline
Biais & RMSE & Biais & RMSE & Biais & RMSE& Biais & RMSE& Biais & RMSE\\
\hline
\hline
 -9e-04&0.0102 & 0&0.0011 & 0&4e-04 & 1e-04&6e-04 & -1e-04&9e-04 \\
 8e-04&0.0064 & 1e-04&2e-04 & -0.0036&0.0557 & 0.0013&0.0085 & 0&0 \\
 1e-04&7e-04 & -1e-04&4e-04 & 0&0 & 0&2e-04 & 0.0026&0.0155 \\
 2e-04&0.0194 & 4e-04&0.0038 & 0.0038&0.0515 & -6e-04&0.0119 & 0&0 \\
 0.0024&0.0281 & -0.0023&0.0099 & 0&0 & 0&0 & -0.0012&0.0077 \\
0&0 & 0&0 & 0&0 & 0&0 & -0.0081&0.0323 \\
   0&0 & 0&0 & 0.002&0.0553 & -0.0039&0.0212 & 0&0 \\
   \hline
\end{tabular}
   \caption{\emph{Simulation scenario 1} : Biais and RMSE for $\alpha^{qq'}_{kk'}$.}  
\label{tab : simu 1 alpha}
\end{table}

\begin{figure}
\centering
  \begin{subfigure}[b]{0.44\textwidth}
\centering
   \includegraphics[ width = \textwidth]{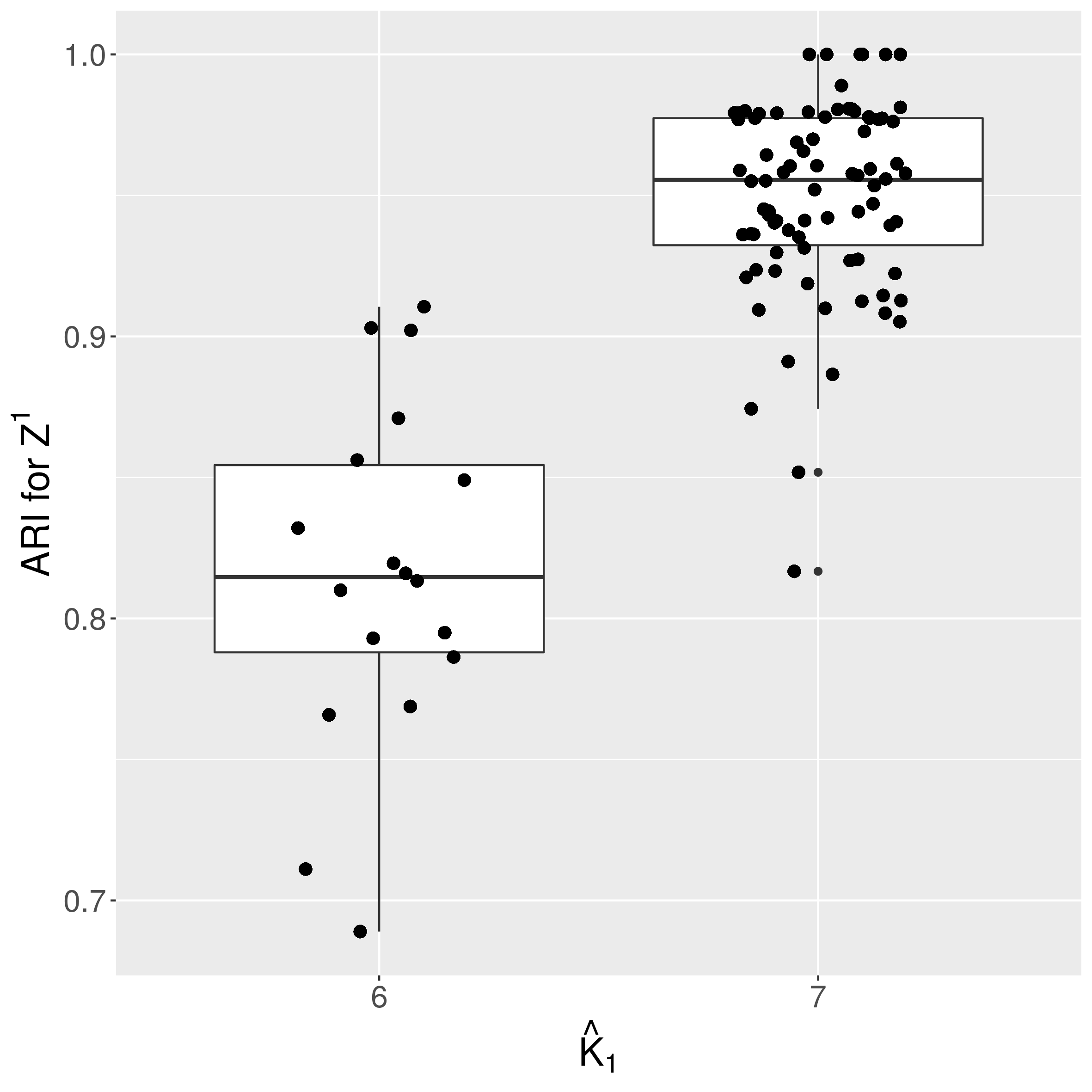}
 \caption{}
   \label{fig:simu1 ARI}
\end{subfigure}
  \begin{subfigure}[b]{0.44\textwidth}
\centering
   \includegraphics[width = \textwidth]{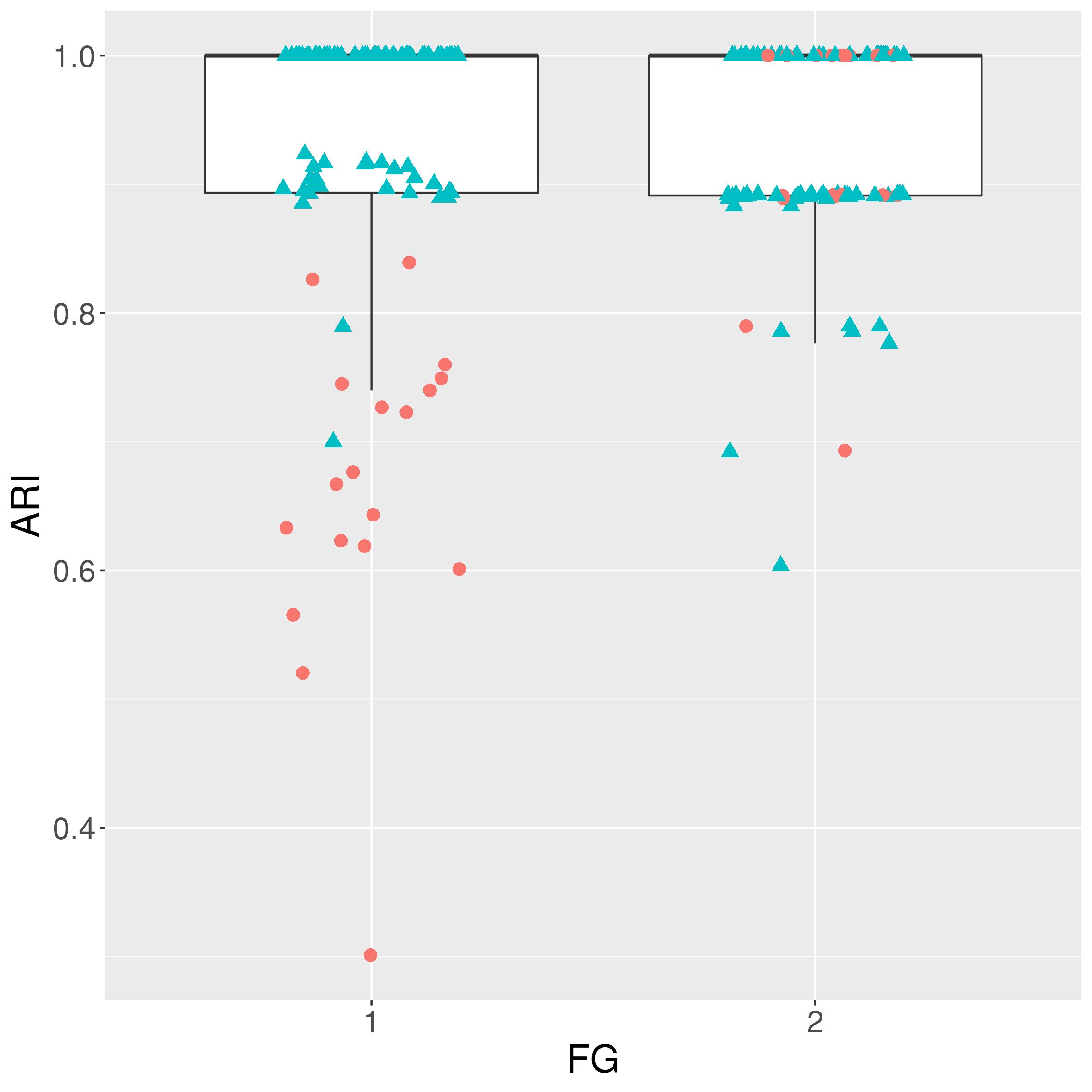}
\caption{}
   \label{fig:simu2 ARI}
\end{subfigure}
   \caption{ (a) : \emph{Simulation scenario 1} : distribution of the ARI for  $q=1$ when $\hat K_1= 6$ (under-estimation) on left and $\hat K_1=7$  (exact estimation) on right.  (b) \emph{Simulation scenario 2} : Boxplots of the ARI for $\hat{\bZ}^1$  (on left) and $\hat{\bZ}^2$ (on right) with  $\hat{K}_1 =2$ (circle) and   $\hat{K}_1 =3$ (triangle).    }
 \end{figure}

Among the $100$ simulated datasets, the true numbers of clusters are   exactly  recovered for $73$ datasets as can be seen in Table \ref{tab : estim K Dattilo}.  Three  other estimated configurations are  observed as detailed in Table \ref{tab : estim K Dattilo}, each of them corresponding to $ \hat{K}_q = K_q-1$.
A scrutiny shows that cases where $K_3$ is estimated to $1$ instead of $2$ correspond to datasets where one simulated cluster (meaning the simulated $\bZ^{3})$   is reduced to  $1$ or $2$ individuals.\\
We measure the ability of the procedure to recover the clusters by computing  the Adjusted Rand Index \citep[ARI by ][]{hubert85}  between $\hat{\bZ}^1$ and the simulated $\bZ ^1$.  ARI  compares two clusterings with a correction for chance. It is close to $0$ when the two clusterings are independent and equals $1$ when the clusterings are identical (up to label switching). 
The boxplots of the   ARI's are plotted in Figure \ref{fig:simu1 ARI}.  All the ARI are high ($>0.7)$, even when $\hat K_1 = 6$, which means that the clusters are always globally recovered  and the  structure of the multipartite network is well discovered.
\\
We also investigate   the  quality of the parameters  estimation. For the $73$ datasets such that $\hat{\bK} =  (7,2,2,1)$,      we compare the estimated parameters $\hat \alpha$  to the simulation parameters (after a relabeling of the clusters to match the true clustering, if required). The results in terms of biais and Root Mean Square Error (RMSE) are reported in Table \ref{tab : simu 1 alpha}.  
 We observe that the parameters are estimated without noticeable biases and with small  RMSE.  In particular, the null  $\alpha_{kk'}^{qq'}  $'s  are always recoverered.

\textbf{Results for Scenario 2}
The same analysis is done for Scenario 2.  $\hat \bK = \bK = (3,2)$   for $82$ simulated datasets over $100$.  The unique alternative observed configuration is $\hat \bK= (2,2)$.
\\Here again, in order to assess the ability of the method to recover the clusters, we compute the ARI for $\bZ^1$ and $\bZ^2$. These quantities are reported in Figure \ref{fig:simu2 ARI} for the two functional groups (left au right) and for $\hat{K}_1 = 3$ (triangles) and $\hat{K}_2 = 2$ (circles). As can be noticed, the clusterings are well  recovered when the number of clusters are exactly recovered (triangle), with a noticeable number of cases where the clusters are exactly recovered (ARI$ = 1$). As expected, we observe a lower ARI when $\hat K_1 = 2$ (circle), but with globally high values.

We also pay attention to the estimation of the parameters of connection $\alpha_{kk'}^{qq'}$.  Biais and RMSE are reported in Table \ref{tab : simu 2 alpha}. Once again, the biais and RMSE are small.

\begin{table}[ht]
\centering
\small
\begin{tabular}{|ll|ll|ll||ll|ll|}
  \hline
\multicolumn{6}{|c||}{$\alpha^{11}$} &  \multicolumn{4}{c|}{$\alpha^{12}$} \\
\hline
\hline
\multicolumn{2}{|c|}{$\alpha_{k1}^{11}$} &  \multicolumn{2}{c|}{$\alpha^{11}_{k2}$} &  \multicolumn{2}{c||}{$\alpha^{11}_{k3}$} &   \multicolumn{2}{c|}{$\alpha^{12}_{k1}$} &   \multicolumn{2}{c|}{$\alpha^{12}_{k2}$}\\
\hline
  \hline
Biais & RMSE & Biais & RMSE & Biais & RMSE& Biais & RMSE& Biais & RMSE\\
\hline
\hline
-0.0039&0.0211 & -1e-04&0.0299 & -5e-04&0.0271 & -0.0043&0.0263 & -0.0192&0.0552 \\
 0.0021&0.0342 & -0.0014&0.045 & 6e-04&0.0264 & -0.007&0.0348 & -0.026&0.0378 \\
-0.0041&0.0636 & -0.0032&0.0479 & 0.003&0.0826 & -0.0179&0.0377 & -0.0202&0.0534 \\
   \hline
\end{tabular}
   \caption{\emph{Simulation scenario 2} : Biais and RMSE for $\alpha^{qq'}_{kk'}$.}  
\label{tab : simu 2 alpha}
\end{table}

%% file: annexe_detailsVEM.tex
\newpage
\allowdisplaybreaks
\section{Variational EM for MBM: details} \label{app:VEM}


%
%
%


\vspace{1em}

The variational version of the EM algorithm maximizes a lower bound of the observed likelihood.
More precisely,  let  $\Rapprox$ be any probability  distribution on $\bZ$, we define  $\Ical_{\theta}(\Rapprox) $  as:

\begin{eqnarray}
\Ical_{\theta}(\Rapprox) & =&   \log   \ell(\bX ; \theta)  - \KL[\Rapprox, \Prob(\cdot | \bX; \theta)]\label{eq:I1}\\
 &=&   \Expec_{\Rapprox}\left[\log \ell_c(\bX, \bZ; \theta)\right] + \mathcal{H}(\Rapprox)  \label{eq:I2}\\
 &\leq&  \log \ell(\bX ; \theta) \label{eq:I3}
  \end{eqnarray}
where $\mathbf{KL}$ is the Kullback-Leibler divergence and $\mathcal{H}(\Rapprox)$  is the entropy of $\Rapprox$. The inequality in  \eqref{eq:I3} derives from the positivity of the $\KL$ divergence. The equality $\Ical_{\theta}(\Rapprox)  = \log \ell(\bX; \theta)$ holds iff  $\Rapprox(\bZ) = $ $ \Prob(\bZ | \bX; \theta)$ which results in the classical identity of the EM algorithm \citep{dempster77}. 

The principle of the VEM algorithm is to choose $\Rapprox$ in a family of distributions   $\Pcal$ parametrized
by $\boldsymbol{\tau}$,
such that the conditional expectation $  \Expec_{\Rapprox}\left[\log \ell_c(\bX, \bZ; \theta)\right] $ can be computed explicitly.
Iteration $(t)$ of VEM consists in the two following steps:
\begin{itemize}
\item[$\bullet$]\textbf{M Step}
\begin{eqnarray*}
\theta^{(t)} &=& \argmax_{\theta}   \Ical_{\theta}(\mathcal{R}_{\btau^{(t-1)},\bX})\\
&  =&  \argmax_{\theta}   \Expec_{\Rapprox}\left[\log \ell_c(\bX, \bZ; \theta)\right]
\end{eqnarray*} 
\item[$\bullet$]\textbf{VE Step}
\begin{eqnarray*}
\btau^{(t)} &=&  \argmax_{\boldsymbol{\tau}}    \Expec_{\Rapprox}\left[\log \ell_c(\bX, \bZ; \theta)\right] + \mathcal{H}(\Rapprox)\\
&=&  \argmin_{\boldsymbol{\tau}}\KL[\Rapprox, \Prob(\cdot | \bX; \theta^{(t)})] \,.
\end{eqnarray*}
\end{itemize}

 The variational EM generates a sequence $(\theta^{(t)},\btau^{(t)})_{t \geq 0}$ increasing the lower bound $\Ical_{\theta}(\Rapprox)$ of the likelihood $\log \ell(\Xall ; \theta)$.

 \vspace{0em}

\paragraph{Choice of $\Pcal$}

The key point of the procedure is the choice of $\Pcal$, making the calculus tractable but rich enough to be a good approximation of the true conditional distribution   $\Prob(\bZ | \bX; \theta)$. Following  \cite{Govaert2008}  and \cite{Daudinetal2008}, we adopt the mean-field strategy   \citep{jaakkola00} and chose  $\mathcal{P}$ as:
\begin{equation}\label{eq:R}
 \nonumber
\mathcal{P} = \left\{ \mathcal{R}_{ \btau} | \mathcal{R}_{ \btau}(\Zall) = \prod_{q=1}^Q \prod_{i=1}^{n_q} h_{K_q}(Z^q_i; \btau^q_i)\right \}\,,
 \end{equation}
 where $h_{K_q}(\cdot; \xi)$ is the density of a   $1$ trial - multinomial distribution of parameter  $\xi \in \mathcal{T}_{K_q}$,
 i.e. $\mathcal{R}_{\Xall,\btau}$  is such that  the latent variables $\bZ$ are   independent  and $\Prob_{\Rapprox}(Z^q_i=k) = \tau_{ik}^q$
with
$$\sum_{k=1}^{K_q} \tau^q_{ik}=1,  \quad \quad   \forall q \in \{1,\ldots,Q\},  \forall i \in \{1,\ldots,n_q\}. $$
 From this particular choice of $\Pcal$, we derive
 the VE-step and M-step for the MBM in what follows.
 \vspace{1em}



%
%

\paragraph{Explicit expression for $\mathcal{I}_{\theta}(\Rapprox)$}

  Using the expression of the complete log-likelihood given in Equation \eqref{eq:lik}, we obtain:

\begin{equation}\label{eq:lik2}
\begin{array}{cl}
&\mathcal{I}_{\theta}(\Rapprox) =  
 -   \sum_{q,i,k} \tau_{ik}^q  \log \tau_{ik}^q + \sum_{q,i,k} \tau_{ik}^q \log( \pi^q_k)\\
 &+  \sum_{(q,q')} \sum_{(i,i') }  \sum_{(k,k')}\mathbb{E}\left[  \ind_{Z^{q}_i=k}\ind_{Z^{q'}_{i'}=k'} \right] f_{qq'}(X_{ii'}^{qq'},\alpha_{kk'}^{qq'})
    \end{array}
\end{equation}

\begin{equation}\label{eq:b2}
  f_{qq'}(X_{ii'}^{qq'},\alpha_{kk'}^{qq'}) = \left\{
\begin{array}{ll}
 X^{qq'}_{ii'} \log(\alpha^{qq'}_{kk'}) + (1- X^{qq'}_{ii'}) \log (1-\alpha^{qq'}_{kk'})\ & \mbox{for binary networks}\\
-\alpha^{qq'}_{kk'} +  X^{qq'}_{ii'} \log(\alpha^{qq'}_{kk'}) - \log  ( X^{qq'}_{ii'} !)  & \mbox{for weigthed Poisson networks.}\\
\log \phi(X^{qq'}_{ii'} , \mu^{qq'}_{kk'}, \sigma^{qq'}_{kk'})& \mbox{for Gaussian networks.}
\end{array}
\right.
  \end{equation}

%

  \vspace{1,5em}

\noindent $\mathbb{E}\left[  \ind_{Z^{q}_i=k}\ind_{Z^{q'}_{i'}=k'} \right]$ has to be carefully calculated, when $i=i'$.
 To that purpose, let us introduce the following notations :
\begin{itemize}
\item[$\bullet$] $\forall q$,  $\Ecal_q  = \left\{q'  \in \{1,\ldots,Q\} \;  |  \; q' \neq q   \mbox{ and }  (q,q') \in \Ecal\right\}. $
 $\Ecal_q$ is the set of incidence matrices involving the functional group $q$.
 \item[$\bullet$] $\forall (q,q') \in \Ecal, \forall i  \in [\!1, n_q\}$ we define :
$$ \Scal^{qq}_i=\left \{i' \in \{1,\ldots, n_{q}\}\;  |  \; i'\not=i, \ (i,i') \in \Scal^{qq} \right\} \,.$$
\end{itemize}

\vspace{1em}
 
\noindent Using these notations we detail   the expression of  $\mathcal{I}_{\theta}(\Rapprox)$.
\begin{eqnarray*}
&&\sum_{(q,q')} \sum_{(i,i') }  \sum_{(k,k')}\mathbb{E}\left[  \ind_{Z^{q}_i=k}\ind_{Z^{q'}_{i'}=k'} \right] f_{qq'}(X_{ii'}^{qq'},\alpha_{kk'}^{qq'})  \\
&=&  \sum_{q}\sum_{q' \in \Ecal_q} \sum_{(i,i') }  \sum_{(k,k')} \tau_{ik}^q \tau_{i'k'}^{q'} f_{qq'}(X_{ii'}^{qq'},\alpha_{kk'}^{qq'})  +\ind_{(q,q)\in{\mathcal E}}  \sum_{(i,i') }  \sum_{(k,k')}  \mathbb{E}\left[  \ind_{Z^{q}_i=k}\ind_{Z^{q}_{i'}=k'} \right] f_{qq'}(X_{ii'}^{qq},\alpha_{kk'}^{qq})\\
&=&  \sum_{q}\sum_{q' \in \Ecal_q} \sum_{(i,i') }  \sum_{(k,k')} \tau_{ik}^q \tau_{i'k'}^{q'} f_{qq'}(X_{ii'}^{qq'},\alpha_{kk'}^{qq'})   +   \ind_{(q,q)\in{\mathcal E}}  \sum_{i}  \sum_{i' \in \Scal^{qq}_i } \sum_{(k,k')}  \mathbb{E}\left[  \ind_{Z^{q}_i=k}\ind_{Z^{q}_{i'}=k'} \right] f_{qq}(X_{ii'}^{qq},\alpha_{kk'}^{qq})\\
&& + \ind_{(q,q)\in{\mathcal E}}  \sum_{i}   \sum_{(k,k')}  \mathbb{E} \underbrace{\left[ \ind_{Z^{q}_i=k}\ind_{Z^{q}_{i}=k'}\right]}_{=0 \mbox{ if } k \neq k'}   f_{qq}(X_{ii}^{qq},\alpha_{kk'}^{qq})\\
&=&   \sum_{q}\sum_{q' \in \Ecal_q} \sum_{(i,i') }  \sum_{(k,k')} \tau_{ik}^q \tau_{i'k'}^{q'} f_{qq'}(X_{ii'}^{qq'},\alpha_{kk'}^{qq'})  +\ind_{(q,q)\in{\mathcal E}}  \sum_{i}  \sum_{i' \in \Scal^{qq}_i } \sum_{(k,k')}  \tau_{ik}^q \tau_{i'k'}^{q} f_{qq}(X_{ii'}^{qq},\alpha_{kk'}^{qq})\\
&& + \ind_{(q,q)\in{\mathcal E}}  \sum_{i \; | \; (i,i) \in \Scal^{qq}}   \sum_{k}  \mathbb{E}  \underbrace{\left[ \ind_{Z^{q}_i=k}^2\right] }_{= \ind_{Z^{q}_i=k}}f_{qq}(X_{ii}^{qq},\alpha_{kk}^{qq})\,.
\end{eqnarray*}
As a consequence,
we get:
\begin{equation}\label{eq:I}
\begin{array}{ccl}
\mathcal{I}_{\theta}(\Rapprox) &=&  
 -   \sum_{q,i,k} \tau_{ik}^q  \log \tau_{ik}^q + \sum_{q,i,k} \tau_{ik}^q \log( \pi^q_k)\\
&& +     \sum_{q}\sum_{q' \in \Ecal_q} \sum_{(i,i') }  \sum_{(k,k')} \tau_{ik}^q \tau_{i'k'}^{q'} f_{qq'}(X_{ii'}^{qq'},\alpha_{kk'}^{qq'}) \\
&& +\ind_{(q,q)\in{\mathcal E}}  \sum_{i}  \sum_{i' \in \Scal^{qq}_i } \sum_{(k,k')}   \tau_{ik}^q \tau_{i'k'}^{q}   f_{qq}(X_{ii'}^{qq},\alpha_{kk'}^{qq})\\
&& + \ind_{(q,q)\in{\mathcal E}}  \sum_{i \; | \; (i,i) \in \Scal^{qq}}   \sum_{k}  \tau_{ik}^q   f_{qq}(X_{ii}^{qq},\alpha_{kk}^{qq})\,.
\end{array}
\end{equation}

\paragraph{Optimization of $\mathcal{I}_{\theta}(\Rapprox)$ with respect to $\btau$, ($\theta$ being fixed)} $\:$ \\

\noindent  For a fixed $\theta$, we need to find $\btau$ such  that
$ \forall q \in \{1,\ldots, Q\}$, $\forall k  \in \{1,\ldots,  K_q\}$, $ \forall i \in \{1,\ldots,  n_q\}$:
\begin{equation}
\frac{\partial}{\partial \tau^q_{ik}} \left[\mathcal{I}_{\theta}(\Rapprox) + \sum_{q'=1}^Q\sum_{j=1}^{n_{q'}}\lambda^{q'}_{j}\left( \sum_{k'=1}^{K_q}\tau_{jk'}-1\right)\right] = 0 \label{eq:derivtau}
\end{equation}
where $(\lambda^{q'}_j)_{1\le q'\le Q, 1\le j\le n_{q'}}$ are the Lagrange multipliers.
Combining Equations \eqref{eq:lik2} and \eqref{eq:I}, we get:
\begin{eqnarray}\label{eq:syst2}
0 &=&  -(1 + \log \tau_{ik}^q)+  \log\pi_k^q  +   \left[ \sum_{q' \in \mathcal{E}_q}  \sum_{i'=1}^{n_{q'}}\sum_{k' = 1}^{K_q'} f_{qq'}(X_{ii'}^{qq'},\alpha_{kk'}^{qq'}) \tau_{i'k'}^{q'} \right]  \nonumber \\
&&  +   \ind_{(q,q)\in{\mathcal E}}  \sum_{j \in \Scal^{qq}_i } \sum_{k'=1}^{K_q} f_{qq}(X_{ij}^{qq},\alpha_{kk'}^{qq}) \tau_{jk'}^{q}\\
&& +   \ind_{(q,q)\in{\mathcal E}}     \ind_{(i,i) \in \Scal^{qq}}   f_{qq}(X_{ii}^{qq},\alpha_{kk}^{qq}) \nonumber  \\
  &&+ \lambda_{i}^q  \,.  \nonumber
\end{eqnarray}


\noindent This system has no explicit solution but can be solved numerically using a fixed point strategy  as in \cite{Daudinetal2008}.

\paragraph{Optimization of $\mathcal{I}_{\theta}(\Rapprox)$ with respect to $\theta$, $\btau$ being fixed.}   $\;$ \\
\noindent We have to compute the derivatives of $\mathcal{I}_{\theta}(\Rapprox)$ with respect to   $\theta$,  the variational parameters $\btau$ being fixed. We thus obtain:  $\forall (q,q') \in \Ecal, \forall (k,k') \in \{1,\ldots, K_q\} \times\{1,\ldots, K_{q'}\}$:
$$
\alpha^{qq'}_{kk'} = \frac{\sum_{(i,i')\in \Scal^{qq'}} X^{qq'}_{ii'}\tau^q_{ik}\tau^{q'}_{i'k'}}{\sum_{(i,i')\in \Scal^{qq'}}   \tau^q_{ik}\tau^{q'}_{i'k'}}
 $$
 and $\forall q \in \{1,\ldots, Q\},  \forall k=\{1,\ldots, K_q\}$:
 $$
\pi_{k}^q = \frac{1}{n_q} \sum_{i=1}^{n_q}\tau^q_{ik}\,.
  $$

\paragraph{Clustering and initializations}$\;$\\
 We denote by $ \widehat \theta$ and $\widehat \btau$ the resulting estimates. The estimated clustering is the maximum a posteriori (MAP) estimate:  $ \forall q \in \{1,\ldots,Q\},  \forall  i \in\{1,\ldots,n_q\}$,
$$\hat{Z}_i^q = \argmax_{k \in \{1,\ldots,K_q\}}\; \widehat{\tau}_{ik}^q . $$

%
%

%% file: annexe_ICL.tex
\newpage
\section{Derivation of the ICL criterion}\label{app:proofICL}

As exposed in Section \ref{subsec:modselec} of the paper, we resort to the ICL criterion to perform model selection.  The ICL is an asymptotic approximation of the integrated marginal complete likelihood.  We supply here the details of the calculations.

\paragraph{Explicit expression of the marginal  complete likelihood}

The ICL being a based on a Bayesian model selection criterion, we first set the following prior distribution: 
\begin{equation}\label{eq:prior_A}
\alpha^{qq'}_{kk'}  \sim \mathcal{B}(a,a) \quad \mbox{and } \quad  (\pi^q_{1}, \dots \pi^q_{K_q}) \sim \mathcal{D}ir(b, \dots,b).
\end{equation}
where $\mathcal{B}(a,a) $ denotes the Beta distribution and   $\mathcal{D}ir(b, \dots,b)$ is the Dirichlet distribution. 
 By definition,  $$\log m_c(\bX,\bZ;  \M)   =  \log  \int \ell_c(\Xall,\bZ; \theta_{\bK}) \prior(\theta_{\bK};  \M)d \theta_{\bK}\,.$$
The prior on $\theta$ being such that $\prior(\theta) =\prior(\balpha) \prior(\bpi)$ we obtain:
\begin{eqnarray*}
\log m_c(\bX,\bZ;  \M) &=&  \log  \int f(\Xall| \bZ; \balpha ) \prior(\balpha) d\balpha  +    \log  \int f(\bZ; \bpi )\prior(\bpi) d\bpi\,.
\end{eqnarray*}
Taking advantage of the conditional independences in the model defined by Equations  \eqref{eq:mod2} and \eqref{eq:mod1} combined with the independence of the parameters in the prior distribution,
we can decompose $\log m_c$ into the following sum:
\begin{eqnarray*}
\log m_c(\bX,\bZ;  \M)&=& \sum_{(q,q')\in\Ecal}\log  \int f(\bX^{qq'} | \bZ^q, \bZ^{q'}; (\balpha^{qq'}))\prior(\balpha^{qq'})  d\balpha^{qq'}\\
&& +  \sum_{q=1}^Q  \log   \int f(\bZ^q; \bpi^q) \prior(\bpi^q) d \bpi^{q}\,.\\
\end{eqnarray*}
Using the fact that $  f(\bZ^q; \bpi^q) = \prod_{k=1}^{K_q} (\pi_k^q)^{N_k^q}$ with
\begin{equation}\label{eq_Nkq}
 N_k^q = \sum_{i=1}^{n_q}
\ind_{Z_i^q = k}
\end{equation}
and  the conjugacy of the Dirichlet prior distribution, we easily deduce the following formula:
$$ \int f(\bZ^q; \bpi^q) \prior(\bpi^q) d \bpi^{q}  = \frac{ \Gamma(bK_q)}{\Gamma(b)^{K_q}}  \frac{ \prod_{k=1}^{K_q} \Gamma(N_k^q+b)}{\Gamma(n_q+bK_q)} $$
where $\Gamma$ is the Gamma function.

\vspace{1em}

\noindent Now,  we can reformulate $ f(\bX^{qq'} | \bZ^q, \bZ^{q'};  \balpha^{qq'} )$  as:
\begin{eqnarray*}
 f(\bX^{qq'} | \bZ^q, \bZ^{q'}; \balpha^{qq'} ) &=& \prod_{(i,i',k,k')}   (\alpha_{kk'}^{qq'})^{X_{ii'}^{qq'} \ind_{Z_i^q=k} \ind_{Z_{i'}^{q'}=k}}  (1- \alpha_{kk'}^{qq'})^{(1-X_{ii'}^{qq'}) \ind_{Z_i^q=k} \ind_{Z_{i'}^{q'}=k}}\\
 &=&  \prod_{k,k'=1}^{K_q, K_{q'}}   (\alpha_{kk'}^{qq'})^{S^{qq'}_{kk'}}   (1- \alpha_{kk'}^{qq'})^{N^{qq'}_{kk'} - S^{qq'}_{kk'}}
\end{eqnarray*}
with
\begin{equation}\label{eq_S}
\begin{array}{ccl}
 S^{qq'}_{kk'} &=&  \sum_{(i,i') \in \Scal^{qq'}} X_{ii'}^{qq'} \ind_{Z_i^q=k} \ind_{Z_{i'}^{q'}=k} \\
  N^{qq'}_{kk'} &=&  \sum_{(i,i') \in \Scal^{qq'}}   \ind_{Z_i^q=k} \ind_{Z_{i'}^{q'}=k}.
  \end{array}
 \end{equation}
With the beta prior distribution on each $\alpha_{kk'}^{qq'}$, we get:
\begin{eqnarray*}
&& \int f(\bX^{qq'} | \bZ^q, \bZ^{q'}; (\balpha^{qq'}))\prior(\balpha^{qq'})  d\balpha^{qq'} =  \prod_{k,k'=1}^{K_q,K_{q'}} \frac{\Gamma(2a)}{\Gamma(a)^2} \frac{\Gamma(a+S^{qq'}_{kk'}) \Gamma(a+N^{qq'}_{kk'} - S^{qq'}_{kk'})}{\Gamma(2a+N^{qq'}_{kk'})}\,.
\end{eqnarray*}
Finally, we obtain:
\begin{eqnarray*}
  \log m_c(\bX,\bZ;  \M) &= &  \left( \sum_{(q,q') \in \Ecal} | \Acal^{qq'}| \right) \left(\log \Gamma(2a) - 2 \log \Gamma(a) \right) \\
&& + \sum_{(q,q',k,k')} \log \Gamma(a+S^{qq'}_{kk'})  + \log \Gamma(a+N^{qq'}_{kk'} - S^{qq'}_{kk'})\\
&&-  \sum_{(q,q',k,k')}  \log \Gamma(2a+N^{qq'}_{kk'})
\\
 && +  \sum_{q=1}^Q \log \Gamma(bK_q) - K_q \log(b)-\log \Gamma(n_q+bK_q) \\
 && +   \sum_{q=1}^Q \ \left(\sum_{k=1}^{K_q}  \log \Gamma(N_k^q+b)\right)
\end{eqnarray*}
where $N_k^q$ has been defined in Equation \eqref{eq_Nkq} and $S^{qq'}_{kk'} $ and $N^{qq'}_{kk'}$ in Equation \eqref{eq_S}

\paragraph{Asymptotic approximation}
Using the same arguments as in \cite{Daudinetal2008} and \cite{braul2014}, we obtain the following asymptotic approximation. Assume that $ \forall  q \in \{1,\ldots,Q\}, n_q \rightarrow \infty$, then :
\begin{eqnarray*}
\log m_c(\bX,\bZ;  \M) & =&  \max_{\theta_{\bK}\in \Theta_{\bK}} \log \ell_c(\bX,\bZ ; \theta_{\bK}) - \mbox{pen}(\M)
\end{eqnarray*}
where
\begin{eqnarray*}
\mbox{pen}(\M)&=&\frac{1}{2}  \sum_{q=1}^Q (K_q-1)\log(n_q) \\
&+&\frac{1}{2} \left(\sum_{(q,q') \in \Scal^{qq'}}  |\Acal^{qq'}| \right)\log\left(  \sum_{(q,q') \in \Scal^{qq'}}  | \Scal^{qq'}|\right)\,.
\end{eqnarray*}
The first term comes from the application of the Stirling formula to the Gamma function when approximating $ f(\bZ^q; \bpi^q)$. The second term comes from a BIC approximation  of the part  $ f(\bX^{qq'} | \bZ^q, \bZ^{q'}; \balpha^{qq'} )$. Obviously,  the  parameters of the prior distribution $(a,b)$  disappear from the formula since asymptotically the importance of the prior distribution vanishes.